\documentclass{achemso}
\usepackage{graphicx}% Include figure files
\usepackage{dcolumn}% Align table columns on decimal point
\usepackage{bm}% bold math
\usepackage{amsmath}
\usepackage{amssymb}
\usepackage{latexsym}
\usepackage{epsfig}
\usepackage{amsbsy}
\usepackage{array}
\usepackage{amssymb}
\usepackage{setspace}
\usepackage{bm}
\usepackage{color}
\usepackage{multirow}
\usepackage{placeins}
\usepackage{subcaption}
\usepackage{siunitx}
\usepackage{soul}
\usepackage[table]{xcolor} 
\usepackage[version=3]{mhchem} % Formula subscripts using \ce{}

\newcommand{\onlinecite}[1]{\hspace{-1 ex} \nocite{#1}\citenum{#1}} 

\title{Numerical analysis of a hysteresis model in perovskite solar cells}
\author{Yecheng Zhou}
\affiliation{School of Chemistry, The University of Melbourne, Parkville, VIC, 3010, Australia} 
\email{yecheng.zhou@unimelb.edu.au}
\author{Fuzhi Huang}
\affiliation{Materials Synthesis and Processing, Wuhan University of Technology, Wuhan, 430070, China.} 

\author{Yi-Bing Cheng}
\affiliation{Department of Materials Science and Engineering, Monash University, Victoria 3800, Australia} 

\author{Angus Gray-Weale }
\affiliation{School of Chemistry, The University of Melbourne, Parkville, VIC, 3010, Australia} 
\email{angusg@unimelb.edu.au}

\begin{document}

%\tableofcontents

\begin{abstract}
Previously, we proposed that the polarization and capacitive charge in \ce{CH3NH3PbI3} screens the external electric field that hinders charge transport. We argue here that this screening effect is in significant part responsible for the power conversion characteristics and hysteresis in \ce{CH3NH3PbI3} photovoltaic cells. In this paper, we implement capacitive charge and polarization charge into the numerical model that we have developed for perovskite solar cells. Fields induced by these two charges screen the applied hindering field, promote charge transport, and improve solar cell's performance, especially in solar cells with short diffusion lengths. This is the reason why perovskite solar cells made from simple fabrication methods can achieve high performance. More importantly, with relaxations of capacitive charge and polarization charge, we quantitatively reproduce experimental ``anomalous'' hysteresis J-V curves. This reveals that both polarization relaxation and ions relaxation could contribute to anomalous hysteresis  in perovskite solar cells. 
\end{abstract}
	\maketitle
%%%%%%%%%%%%%%%%%%%%%%%%%%%%%%%%%%%%%%%%%%%%%%%%%%%%%%%%%%%%%%%%%%%%%%%%%%%%%%%%%%%
\section{Introduction}

Perovskite solar cells have achieved power conversion efficiencies (PCEs) up to 22\% in just five years.\cite{Kojima2009,Record} They attract great attention due to their high performance and anomalous hysteresis. It is believed that the large charge carrier diffusion lengths and the compensated field are two key factors for high performances of hybrid perovskite solar cells. In experiment it was observed that the slowly built compensated field contributes to the anomalous hysteresis.\cite{Tress2015,Unger2014} It is believed that the compensated field is induced by ion migration and electronic charge traps.\cite{Tress2015,Unger2014, VanReenen2015} The compensated field works as a screening effect resulting in a high dielectric constant, which has been observed up to 1000 for \ce{CH3NH3PbI3}.\cite{Juarez-Perez2014,Lin2014}

%\textbullet \textbf{The dielectric constant, screen ability, varies with different measurement scan rates. That is due to compensate field. Its slow response is believed to be induced by ion migration or polarization.}

At high frequencies only electronic orbitals are able to respond and become polarized, whereas at low frequencies all of the electronic orbitals, defect charges and ions are able to respond. As a result, at high frequencies low dielectric constants of 6-7 are observed,\cite{Zhou2015,Juarez-Perez2014} compared to dielectric constants approximately 100 times higher at low frequencies. \cite{Juarez-Perez2014} This frequency dependent dielectric constant behaviour is consistent with the compensated field and the extremely slow photo-conductivity response in \ce{CH3NH3PbI3} solar cells.\cite{Gottesman2014} We argue that both compensated field and slow photo-conductivity come from certain slow relaxations. This relaxation screens external fields and increases the dielectric constant. It is widely believed that ion migration is one possible reason for this slow relaxation.\cite{Unger2014,Beilsten-Edmands2015a,Yang2015,Snaith2014,Tress2015,VanReenen2015} But the slow response experiment carried out by Gottesman \textit{et al.} shows two opposite behaviours of decreased/increased photo-conductivity in identically constructed devices. This cannot be explained by ion migration.\cite{Gottesman2014} It needs to be revised that ion migration is not the only origin of hysteresis. VanReenen \textit{et al.} modeled this hysteresis and found the combination of ion migration and electronic traps brings about hysteresis.\cite{VanReenen2015} 

Besides ion migration, polarization is another possible reason for hysteresis.
Beilsten-Edmands \textit{et al.} claimed that there is no ferroelectric nature contribution to hysteresis due to the intrinsic polarization being too small.\cite{Beilsten-Edmands2015a} They treated the polarization at very high frequency ($f\rightarrow \infty$) as its intrinsic polarization, however this is  incorrect. For any ferroelectric polarizations under a very high frequency it should be zero as the electronic orbitals and ions are unable to respond. There is no direct evidence to deny ferroelectric polarization in perovskite solar cells. Additionally, Kutes \textit{et al.} showed a direct observation of ferroelectric polarizations.\cite{Kutes2014} Debate of polarization in hysteresis continues. 

From a theoretical standpoint most research supports the existence of polarization. First-principles studies have shown that the energy barriers for defect migrations are from 0.08 eV to 0.40 eV depending on ion types\cite{Haruyama2015,Azpiroz2015} This energy barrier is low enough to be crossable at room temperature. We also reported the energy barrier for methylammonium ions (\ce{MA+}) reorientation is about 0.01 eV to 0.098 eV, which depends on the initial and final \ce{MA+} orientations and neighbor \ce{MA+} orientations.\cite{Zhou2015} From the base of their energy landscapes, polarization is easier to respond and screen external fields. \ce{MA+} ions are able to be rotated collectively under an applied external field, which then polarises the \ce{CH3NH3PbI3} crystal or thin film. This collective reorientation and polarization combined with capacitive charges screen external hindering field and promote power conversion efficiency. Hence, another slow relaxation should be polarization relaxation. Our argument is in good agreement with Sanchez's experiment that the slow dynamic process depends strongly on the organic cation, \ce{MA+} or \ce{FA+}.\cite{Sanchez2014}  The rotation of \ce{MA+} ions and migration of \ce{I-} ions are systemically discussed by Frost \textit{et al.}.\cite{Frost2014,Frost2016,Weller2015b} They argued that the internal electrical fields associated with polarization contribute to hysteresis in J-V curves.\cite{Frost2014} They also observed that a single cation rotation and anion migration take several picoseconds.\cite{Frost2016,Weller2015b}  Hence, we argue here that polarization and ion migration are both possible to induce screening fields and contribute to hysteresis.

%\subsection{Hypotheses}
%\textbullet \textbf{High dielectric constant and screening effect benefit to high performance}

\begin{figure} 
	\centering
	\includegraphics[width=8.5cm]{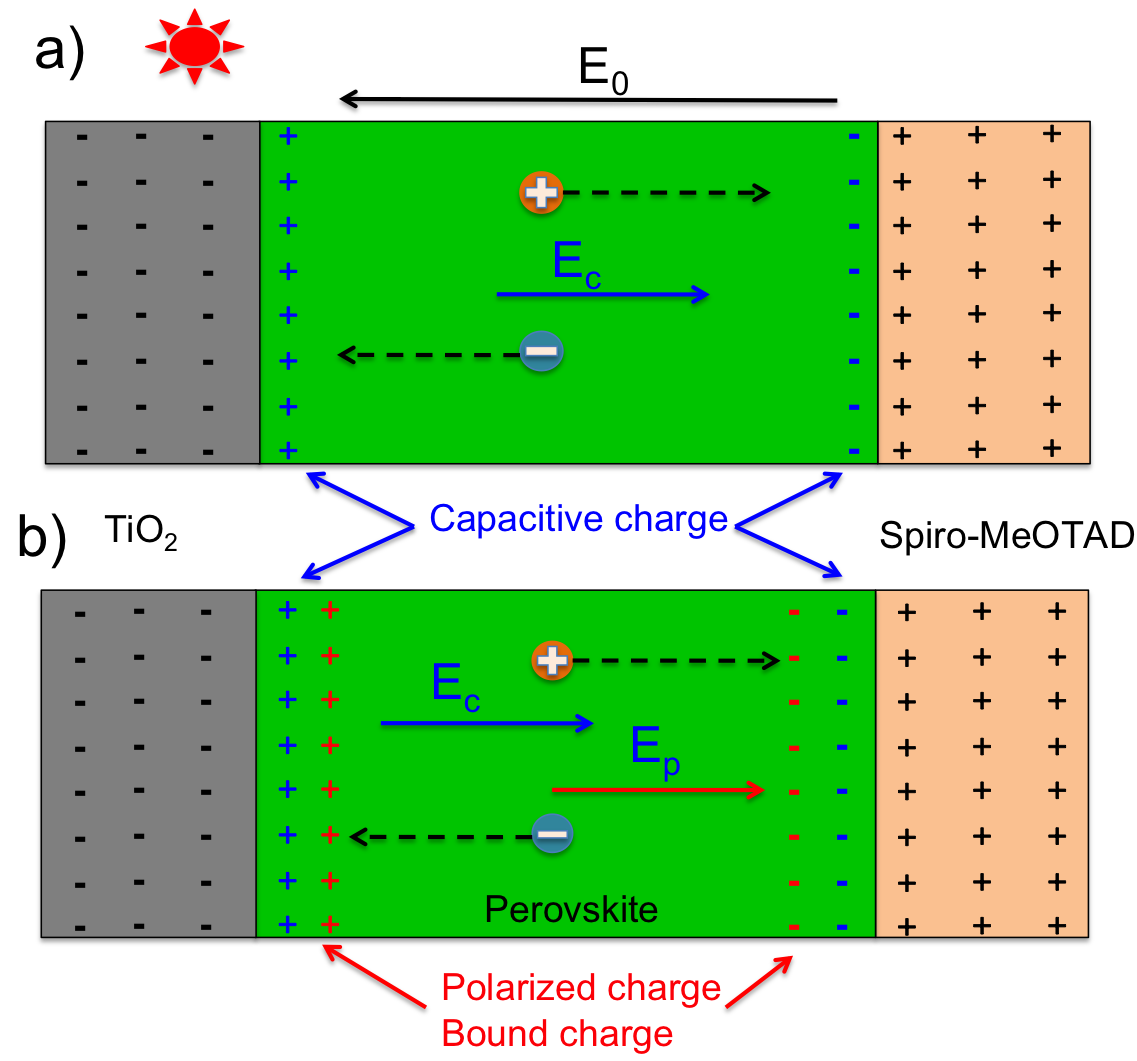}
	\caption{Our proposed mechanism of perovskite solar cells}
	\label{fig:mechanism}
\end{figure}

We propose that the screening charge contains two components: one is the polarization charge  resulted from \ce{MA+} reorientation and inorganic frame, which we name the polarization charge; another is capacitive charges from defects and trapped charges including ions. For normal perovskite solar cells without polarization, photon generated charge carriers accumulate in defects near interfaces, shown as blue charge in Figure \ref{fig:mechanism}. These accumulated charges induce a field that counteracts the applied hindering field. We name the applied field the hindering field because it is opposite to the work current vector. For perovskite solar cells, polarization charges can further counteract the hindering field and promote charge carrier transport. Due to these capacitive and polarization screening effects, high PCEs are expected for hybrid perovskite solar cells. This is the first hypothesis we are going to test. 

The second hypothesis we are going to test is that hysteresis in J-V curves come from these two kinds of slow relaxations. As their relaxations are slow, screening fields fall behind the applied hindering field if the measurement scanning is fast enough. This delay induces hysteresis. We apply numerical simulations to reproduce and explain the anomalous hysteresis effect in perovskite solar cells. Our results show both capacitive charge and polarization charge could contribute to hysteresis effects. Relaxation times of these two charges determine the overall behaviour of scan rate dependent hysteresis.

% % % % % % % % % % % % % % % % % % % % % % % % % % % %
\section{Methods} 
%\subsection{Basic equations and boundary conditions}
Our model is based on the continuity equations and Poisson's equation in one dimension, adapted for perovskite solar cells:\cite{Zhou2016}
\begin{equation}\label{eq:sc} 
\begin{split}
J_n= eD_n\frac{\partial n}{\partial x}+n \mu_n F \\
J_p= -eD_p\frac{\partial p}{\partial x}+p \mu_p F\\
\frac{\partial J_n}{\partial x}=-eG+eR \\ 
\frac{\partial J_p}{\partial x}=eG-eR \\ 
\frac{\partial F}{\partial x}= \frac{p-n}{\varepsilon \varepsilon_0 }
\end{split}
\end{equation}
\noindent
where, $J_n$ and $J_p$ are electron current and hole current respectively; $n$ is electron density and $p$ is hole density; $\mu$ and $D$ are charge carriers mobility and diffusion coefficient respectively; footnote symbols $_n$ and $_p$ mean they belong to electron and hole respectively; $G$ and $R$ are generation rate and recombination rate; $F$ is the external applied electric field. Boundary conditions and parameters are shown in reference \onlinecite{Zhou2016}.

%\subsection{Generation rate, if we publish this as letter, this part should be in SI}

\begin{figure} 
	\centering
	\includegraphics[width=8.5cm]{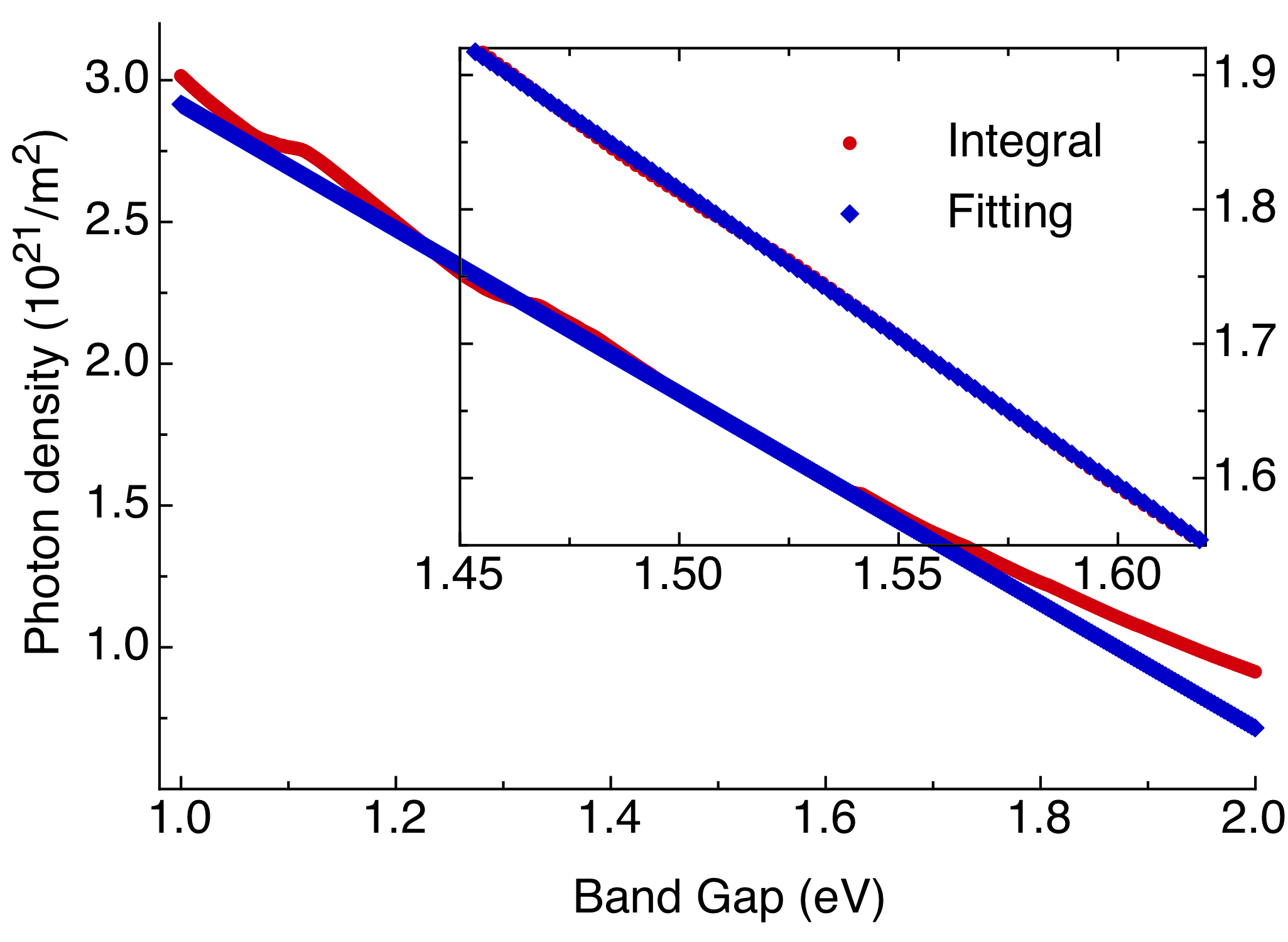} 
	\caption{Integral and fitting incident photon density of AM 1.5 Standard Spectrum. The insert shows the integral and fitting result of between band gaps of 1.45 eV and 1.62 eV. Error in this region is smaller than 2\%.}
	\label{fig:AM1_5}
\end{figure}

Light harvest and charge generation are expressed as $G=IPCE\times N$, where $IPCE$ is Incident Photon-to-Current Efficiency. $N$ is the incident photon density calculated by $\int \frac{I(\lambda)}{hc/\lambda} d\lambda$, where $I(\lambda)$ is incident light density, $h$ is the Plank constant, $c$ is the speed of light and $\lambda$ is the photon's wavelength. According to the Beer-Lambert law, light intensity inside a material decays exponentially from the surface as: $I(\lambda,x)=I(\lambda,0)e^{-\alpha(\lambda) x}$, where $x$ is the incident depth from the surface and $\alpha(\lambda)$ is the absorption coefficient. Therefore, the charge generation rate becomes:
\begin{eqnarray}\label{eq:ag}
	G(x)=\int_0^{\lambda_0}G(\lambda,x) d\lambda =\int_0^{\lambda_0} IPCE(\lambda)\times \frac{I(\lambda,0)\times \alpha(\lambda) \times e^{-\alpha(\lambda) x}}{hc/\lambda} d\lambda
\end{eqnarray}
\noindent
$\lambda_0$ is the absorption edge, corresponding to the bandgap. After a photon is absorbed, an exciton formed by a separated hole and electron pair is generated. The hole and electron attract each other and try to combine. We assume all excitons separate to pairs of free holes and electrons, which means $IPCE(\lambda)=100\%$.\cite{Zhou2016} Then the charge generation rate becomes:
\begin{equation}\label{eq:gr}
	G(x)=\int_0^{\lambda_0} IPCE(\lambda)\times \frac{I(\lambda,0)\times \alpha(\lambda) \times e^{-\alpha(\lambda) x}}{hc/\lambda} d\lambda=\alpha  N_0 e^{-\alpha x}
\end{equation} 
\noindent
where, $N_0 = \int_0^{\lambda_0} \frac{I(\lambda,0)}{hc/\lambda} d\lambda$; $I(\lambda,0)$ is the AM1.5 Standard Solar Spectra. The experiment band gap of \ce{MAPbI3} is in the region from 1.45 eV to 1.70 eV.\cite{Yin2014,Kim2012b,Stoumpos2013,Zhou2014,Yamada2014,Schulz2014}
In our simulations, we use a linear fitting to estimate incident photon density near 1.55 eV. $N_0 =-2.20\times 10 ^{17} \times E_{bgap}$ (eV)+5.12$\times 10 ^{17}$ (cm$^{-2}$), where $E_{bgap}$ is the band gap of the perovskite thin film.
As shown in Figure \ref{fig:AM1_5}, for the band gap in the range of 1.45 eV to 1.62 eV, the accurate integral density and fitted density are almost the same. The incident photon density is calculated to be $1.59\times 10^{17}$ cm$^{-2}$, if the band gap is 1.60 eV. 

%\subsection{Boundary field conditions: Potential decay and field distributions in solar cells, if we publish this as letter, this part should be in SI}
A planar perovskite solar cell has a sandwich structure. Two electrodes clip a compact (\ce{TiO2}) layer, a perovskite layer and a hole transport layer (Spiro-OMeTAD layer). Electrodes are conductors and the potential in a conductor is constant, hence we neglect  their potential drop in this discussion. The most important parts are the clipped compact layer, perovskite layer and hole transport layer. Under light irradiation, charge carriers are generated and flow. Here we describe the resistances of the compact layer, perovskite layer and hole transport layer as $R_t$, $R_p$ and $R_s$ respectively. Based on Ohm's law, the voltage drop in the perovskite layer is $V_p=\frac{V_0 \times R_p}{R_t+R_p+R_s}$,
where $V_0$ is the applied voltage. Two facts refute this  assumption. The first is that Ohm's law only refers to drift current, in which case the current is in the same direction as the field, while current in solar cells is opposite to the applied field. The second is that the diffusion current, which is beyond the Ohm's law and the Drude model description, is larger than the drift current. Hence, it is much more reasonable to consider them as dielectric materials. Then the voltage drop across the perovskite layer is:
  \begin{equation}\label{eq:vp}
  V_p=\frac{V_0\times \epsilon_t \times \epsilon_s \times d_p}{(\epsilon_t\times \epsilon_s \times d_p+\epsilon_t\times \epsilon_p \times d_s+\epsilon_s\times \epsilon_p\times d_t )}= A\times V_0
  \end{equation}
\noindent 
 where $\epsilon$ and $d$ are dielectric constant and thickness of corresponding layers, footnote $_t$ for \ce{TiO2}, $_p$ for perovskite and $_s$ for Spiro-OMeTAD. $A$ is the percentage of applied voltage drop across the perovskite layer, which equals $\frac{ \epsilon_t \times \epsilon_s \times d_p}{(\epsilon_t\times \epsilon_s \times d_p+\epsilon_t\times \epsilon_p \times d_s+\epsilon_s\times \epsilon_p\times d_t )}$. When $\epsilon_t$, $\epsilon_p$ and $\epsilon_s$ are 100,\cite{tio} 1000,\cite{Juarez-Perez2014} and 3,\cite{Snaith2006} and their corresponding layers thickness are 50, 380 and 200 nm,\cite{Zhou2014} then the voltage drop across the perovskite layer is A=0.56\%. For the following simulations, the boundary field at the ends of the perovskite layer are the same, which are $F(x=0)=F(x=d)=\frac{V_p}{d_p}$. 
 
%\subsection{Relaxation of screening capacitive charge and polarization  (or ion migration) charge.}
Defects and traps in semiconductor interfaces are able to charge and discharge as capacitors do. Hence, we name these charges the capacitive charge. In our previous work,\cite{Zhou2015} we also showed that \ce{CH3NH3PbI3} can be polarized by external fields through rotating \ce{MA+} ions and tilting inorganic frame. Due to energy barriers, \ce{MA+} ion needs some time to respond. We name this response the polarization relaxation. We assume that the polarization field relaxes (increases/decreases) exponentially with delay time ($\delta t$): $\delta \textbf{F}[\delta t]= \delta \textbf{F}[\infty] \times (1-e^{- \delta t /\tau_c})$, where $\delta \textbf{F}[\infty]$ is the field difference between the initial field and final field with infinite relaxation time. 

The field in bulk materials within a static external field ($\textbf{F}_0$) is $\textbf{F}=\textbf{F}_0+\textbf{F}_{c}$, where $\textbf{F}_{c}=-S_{c}\textbf{F}_0$ and $S_{c}$ is the screening coefficient for capacitive charges. If there is a polarization field, then the total field becomes $\textbf{F}=\textbf{F}_0+\textbf{F}_{c}+\textbf{F}_{p}$, where $\textbf{F}_{p}=-S_{p}\textbf{F}_0$ and  $S_{p}$ is the screening coefficient due to polarization.
 If the measurement voltage is applied step-by-step, then the field at time $t$ with applied voltage $V$ can be expressed: 
 \begin{equation}
 \textbf{F}[V,t]=\textbf{F}_0[V,t]+\textbf{F}_{c}[V,t]+\textbf{F}_{p}[V,t],
 \end{equation}
 where: $$\textbf{F}_c[V,t]=(\textbf{F}_{c}[V,\infty]-\textbf{F}_c[V-\delta V,t- \delta t])(1-e^{- \delta t /\tau_c})+\textbf{F}_c[V-\delta V,t- \delta t]$$ and 
 $$\textbf{F}_p[V,t]=(\textbf{F}_{p}[V,\infty]-\textbf{F}_p[V-\delta V,t- \delta t])(1-e^{- \delta t /\tau_p})+\textbf{F}_p[V -\delta V,t- \delta t]$$
 in which, $\textbf{F}_{c}[V,\infty]$ and $\textbf{F}_{p}[V,\infty]$ are screening fields under applied voltage $V$ with infinite delay time for capacitive charge and polarization charge, respectively; $\delta t$ is the measurement delay time and $\tau_c$ and $\tau_p$ are relaxation times of capacitive charge and polarization charge respectively; $\textbf{F}[V,t]$ and $\textbf{F}_0[V,t]$ are the total field and the hindering field at time $t$ with applied voltage $V$; $\textbf{F}_c[V-\delta V,t- \delta t]$ and $\textbf{F}_p[V-\delta V,t- \delta t]$ are the field of capacitive charge and the field of polarization charge of last step, respectively. 
 
\section{Results and discussion}
%\subsection{Polarisation induce high dielectric constant promote charge transport and PCEs}

%\textbullet \textbf{List how dielectric constant change the boundary field;}

According to Equation (\ref{eq:vp}), the voltage drop across the perovskite layer is a function of dielectric constant and thickness of each layer. At very slow scan rates (such as 25 mV/s), low frequency dielectric constants are exhibited. 
The low frequency dielectric constants of \ce{TiO2} \cite{tio} and perovskite\cite{Juarez-Perez2014} have values of up to 173 and 1000 respectively. In this case $A$ is 0.5\%. For quick scans, dielectric constants at high frequency are exhibited. The high frequency dielectric constants of \ce{TiO2}, perovskite, Spiro-OMeTAD are 86,\cite{tio} 6,\cite{Juarez-Perez2014} and 3,\cite{Snaith2006} respectively. In this case the potential drop across the perovskite layer is 43\% of the applied voltage.  With a long delay time, screening fields of capacitive charge and polarization charge contribute to the large dielectric constant, and this leads to a small $A$--the percentage of applied voltage drop across the perovskite layer. We propose that the increased dielectric constant of perovskites come from the slow polarization relaxation and ion migration, which has been shown to be possible via DFT calculations.\cite{Zhou2015,Haruyama2015,Azpiroz2015,Weller2015b}

%\textbullet \textbf{Compare simulation curve with Zhou's experiment; PCEs of ideally screened solar cells, upper limit;}
\begin{figure} 
	\centering
	\includegraphics[trim={0mm 0mm 0mm 50mm},clip,width=8.5cm]{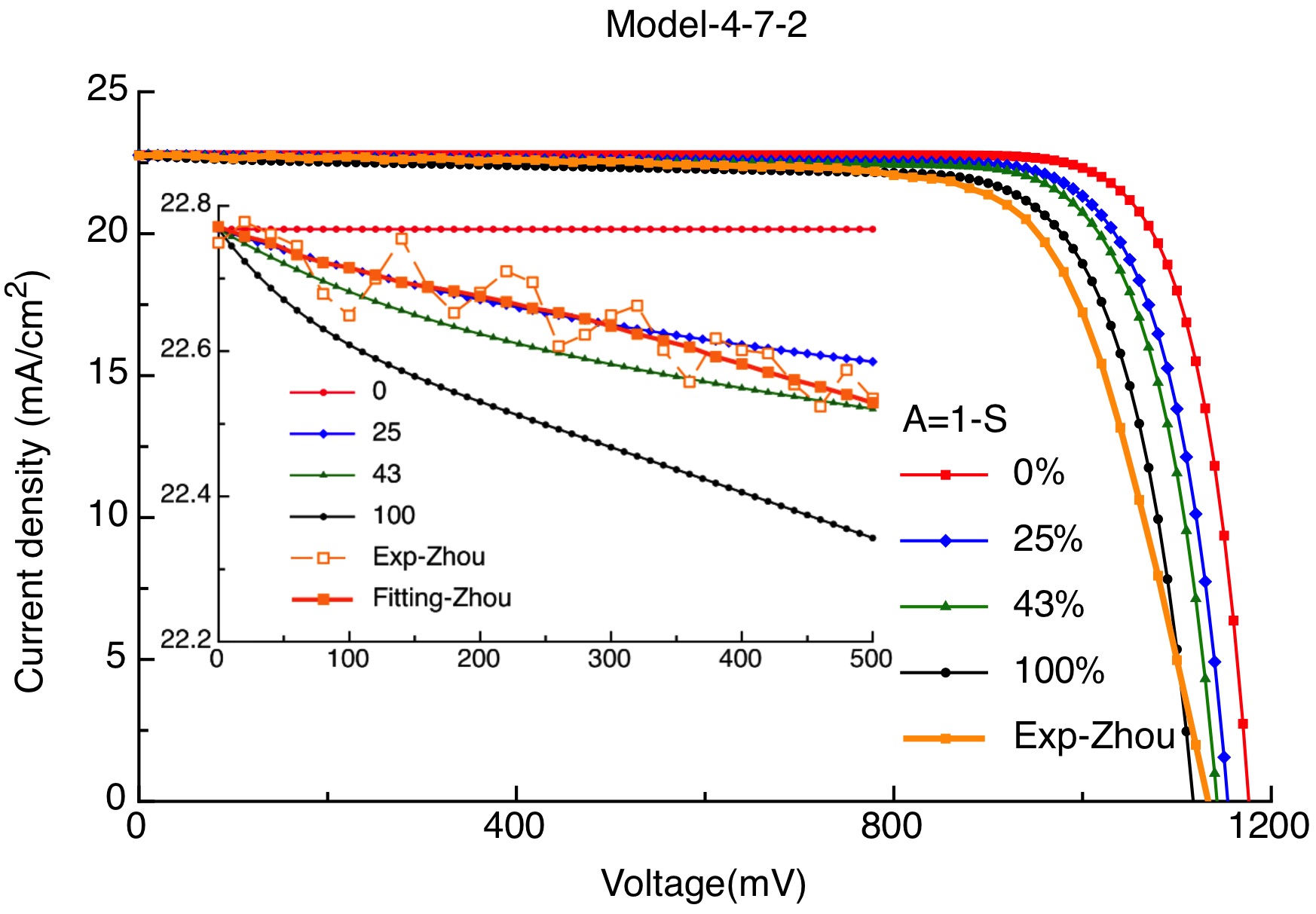}
	\caption{Performances of perovskite solar cells with different $A$ (in Equation \ref{eq:vp}). A 2 nm thick interface with a lifetime of 7 ns is implemented. All of the $J_{sc}$ are \SI{22.77} {mAcm^{-2}}. The measurement delay time is assumed to be infinite.}
	\label{fig:screen}
\end{figure}

\begin{table}[]
	\centering
	\caption{Performance of perovskite solar cells with different values of $A$.}
	\label{tab:Comparison}
	\begin{tabular}{ccccc}
		\hline
		A & $V_{oc}$ (mV)& $J_{sc}$ (mA)& PCEs (\%)& FF \\
		\hline
		0         & 1176   & 22.77    & 22.44    & 0.8380 \\
		25        & 1154  & 22.77    & 21.36    & 0.8131 \\
		43        & 1142  & 22.77    & 20.93    & 0.8044 \\
		100       & 1117      & 22.77       & 19.90    & 0.7821 \\
		Experiment  & 1130      & 22.75       & 19.30    & 0.7507 \\
		\hline 
	\end{tabular}
\end{table}

Based on experimental parameters, the percentage of voltage drop across the perovskite layer is between 0.5\% to 43\%. Here we estimate the solar cell's performance with respect to $A$ with values from 0 to 100\%. The thicknesses of simulated solar cells are 350 nm.\cite{Zhou2014} Diffusion coefficients of the perovskite were set to be \SI{0.017}{ cm^2s^{-1}}(for electrons) and \SI{0.011}{ cm^2s^{-1}}(for holes).\cite{Stranks2013a} The band gap is 1.55 eV.\cite{Zhou2014} Carrier lifetimes in bulk materials without interfaces are assumed to be 736 ns.\cite{Zhou2014} The presence of interfaces decreases charge carrier lifetime which indicates a high recombination rate at the interface. More details have been discussed in reference \onlinecite{Zhou2016}. Thus, we assume the interface recombination region is 2 nm and its charge carriers lifetime is 7 ns, which is determined according to experimental $J_{sc}$. 
In our simulation, for all values of $A$, $J_{sc}$ is \SI{22.77}{ mAcm^{-2}}, which is close to the experimental value of \SI{22.75}{ mAcm^{-2}}. Figure \ref{fig:screen} and Table \ref{tab:Comparison} show experiment and simulation current density--voltage (J-V) curves of solar cells with different $A$. The orange square line is the performance of a solar cell fabricated by Zhou.\cite{Zhou2014}  Our simulated FFs are larger than the experiment value of 75\%. $V_{oc}$ in these simulations are around the experiment value of 1130 mV. 
When $A$ is 100\%, no screening effect is exhibited, and the J-V curve is closest to experiment. However, as shown in the insert of Figure  \ref{fig:screen}, the experiment current decreases between theoretical lines with $A=25\%$ and $A=43\%$ between 0 - \SI{500}{mV} and 22 - \SI{23} { mAcm^{-2}}. Below 300 mV, the blue diamonds line simulated with $A=25\%$ is the closest to Zhou's experiment. The simulated $V_{oc}$, $J_{sc}$, FF and PCE are 1154 mV, 22.77 mA, 81.3\% and 21.36\%, respectively. 

As we proposed that polarization charges can further reduce the hindering field at very low scan rate, then the dielectric constant can be up to 1000.\cite{Juarez-Perez2014} In this case, the voltage drop across the perovskite layer is almost zero, which implies the hindering field has disappeared. With these parameters, our model gives a PCE of 22.4\%.  Therefore. we conclude that by measuring with slow scan rate a high dielectric constant of the perovskite layer results, which reduces the voltage drop across the perovskite layer. Therefore, charge carriers are easier to transfer out, and a higher PCE is achieved. This conclusion is consistent with Sherkar's result that devices having polarization in the plane of devices show high $J_{sc}$ and FF.\cite{Sherkar2015} The difference is that the screening effect in our simulation comes from both polarization and ion migration. It is worth noticing that the polarization comes from both \ce{MA+} ions and the inorganic frame.\cite{Zhou2015}

%\textbullet \textbf{Poor electronic transport ability and screening effect leads to poor performance, agrees with Y-B's experiment.}

\begin{figure} 
	\centering
	\includegraphics[width=8.5cm]{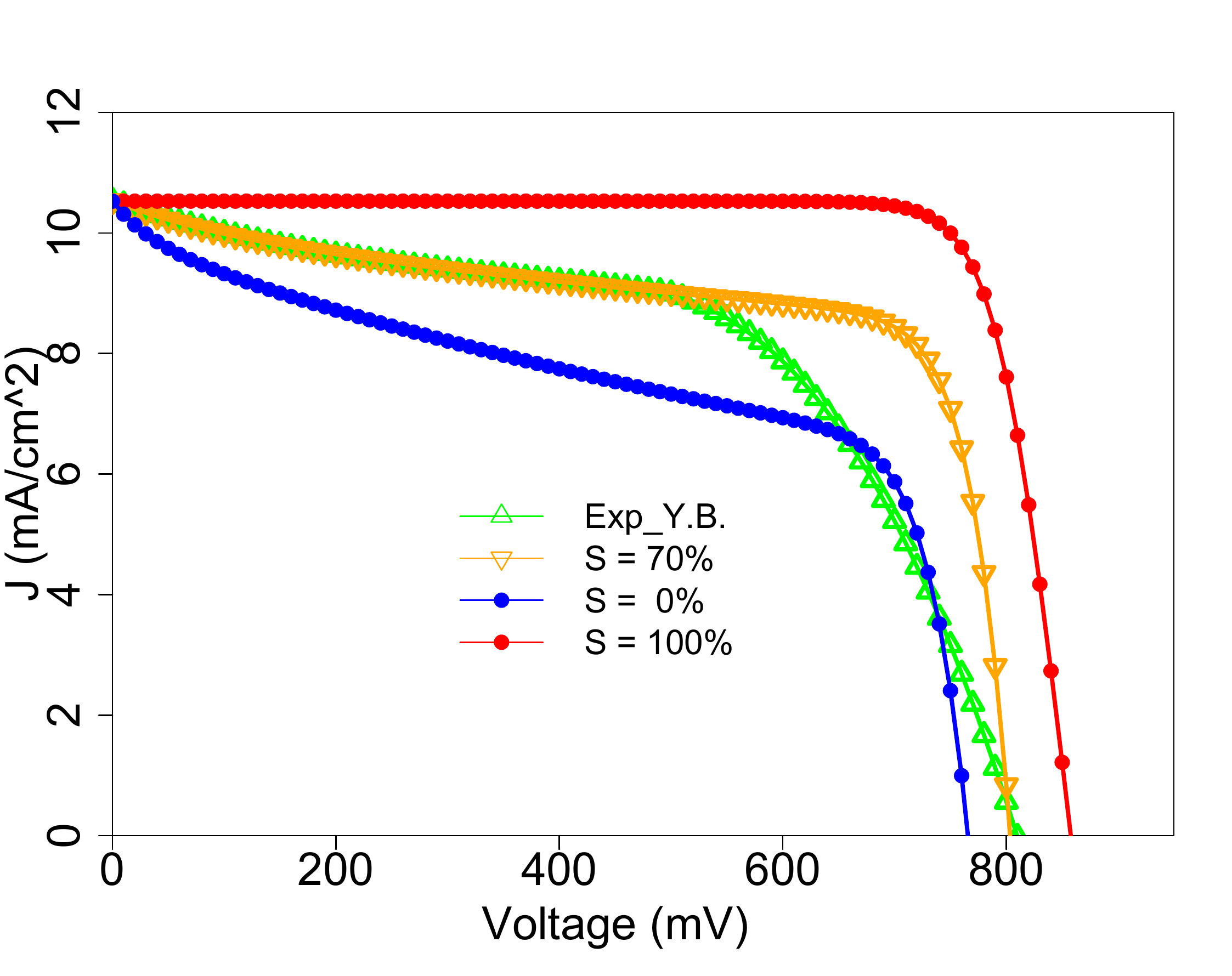}
	\caption{Performance of an unoptimized perovskite solar cell and simulated perovskite solar cells with different screening coefficients, $S$=1-$A$. The thickness of the simulated solar cell is 350 nm. The diffusion coefficients of the perovskite layer were assumed to be \SI{0.017}{ cm^2s^{-1}} and \SI{0.011}{ cm^2s^{-1}} for electrons and holes respectively. The band gap is 1.45 eV. The lifetime in the cell is 57 ns. The working conditions were set to be at 300 K and 1 sun (1.5AM). The charge carrier lifetime in the perovskite layer is 57 ns. The interface region is set as 12 nm thick with an interface charge carrier lifetime of 0.37 ns.}
	\label{fig:yibing}
\end{figure}

The screening improvement is small for solar cells with very high mobilities and long lifetimes. But PCEs of solar cells with poor electron transport ability can be significantly improved through screening effects. A simply made, unoptimized perovskite solar cell shows poor performance as shown by the green triangle in Figure \ref{fig:yibing}. Its performance can be reproduced by our model with short charge carriers lifetime and a thick interface recombination. The simulated $J_{sc}$ is 10.5 \SI{}{mA/cm^2}, in agreement with experiment. 
 For an ideally screened solar cell its current is almost constant before 700 mV. Whereas, for a solar cell without screening effect its current decreases drastically near 0 mV then linearly with voltage until 700 mV. 
The simulated curve with 70\% screen coefficient is in good agreement with experimental curve when the output voltage is lower than 600 mV. The variation at high voltage is due to higher fill factor of our idealized model.
This implies the remaining percentage of the applied voltage in the experimental real solar cell is about 30\%, which is in good agreement with the range of 0.5\% - 43\% calculated from experimental parameters. This is evidence that screening effects are present in perovskite solar cells.
The PCE of the solar cell without screening is 4.3\%, which is improved to 5.9\% if a 70\% screening effect is present. With an ideal screening the PCE would reach 7.5\%. This results in a more than 70\% improvement compared to the unscreened solar cell. The screening effect in solar cells improves PCE, especially in solar cells with poor charge carrier conductivity. Therefore, we draw a conclusion that the screening effect is the reason why simply made perovskite solar cells could achieve high performance.

In a normal solar cell without polarization, screening is also present. But all of screening charges are capacitive charges due to defects and trapped charge carriers. These trapped charges will take part in charge recombination, which in turn reduces its current and PCE. For \ce{MAPbI3}, polarized charges cannot be combined unless polarizations become totally disordered. 
This also gives the benefit of a large current in the device. It is worth noticing that the polarization is not only from \ce{MA+} ions but also from the inorganic frame. For perovskite materials, even though there are no cations with dipoles, the materials can be ferroelectrically polarized as in \ce{BaTiO3}.\cite{batio3} This means perovskite materials, such as \ce{FAPbI3} and \ce{GAPbI3} also have the potential to make high performance solar cells.

In the above simulations, diffusion coefficients were assumed constant. Actually, screening fields also make charge carrier transport easier increasing the diffusion coefficients. Therefore, PCEs of perovskite solar cells can be further improved. 
% % % % % % % % % % % % % % % % % % % % % % % % % % % %

%\subsection{Anomalous hysteresis} 
%\textbullet \textbf{Hysteresis in DSC and silicon solar cells, its timescales.}
 
Capacitive charge is believed to be the main factor for hysteresis in silicon solar cells and DSCs.\cite{Koide2004,Koide2005,Herman2012}  When the measurement scan rate is too fast, capacitive charges are unable to catch up with the changing of scanning field and hysteresis is observed. Usually, more obvious hysteresis is observed with faster scanning.  As charges are able to be trapped and de-trapped in silicon solar cells quickly, an extremely fast scan rate (short delay time) is required to observe hysteresis in silicon solar cells. The measurement delay time is estimated to be around 1 ms.\cite{Herman2012} Charge in DSCs moves slower and the trap and de-trap process takes longer. The relaxation time in DSCs can be up to 100 ms. The required scanning speed to observe hysteresis can be down to \SI{100}{ mVs^{-1}}.\cite{Koide2004,Koide2005} 
 
 %\textbullet \textbf{Normal and anomalous hysteresis in perovskite solar cells.}
  
\begin{figure} 
	\centering
	\includegraphics[width=8.5cm]{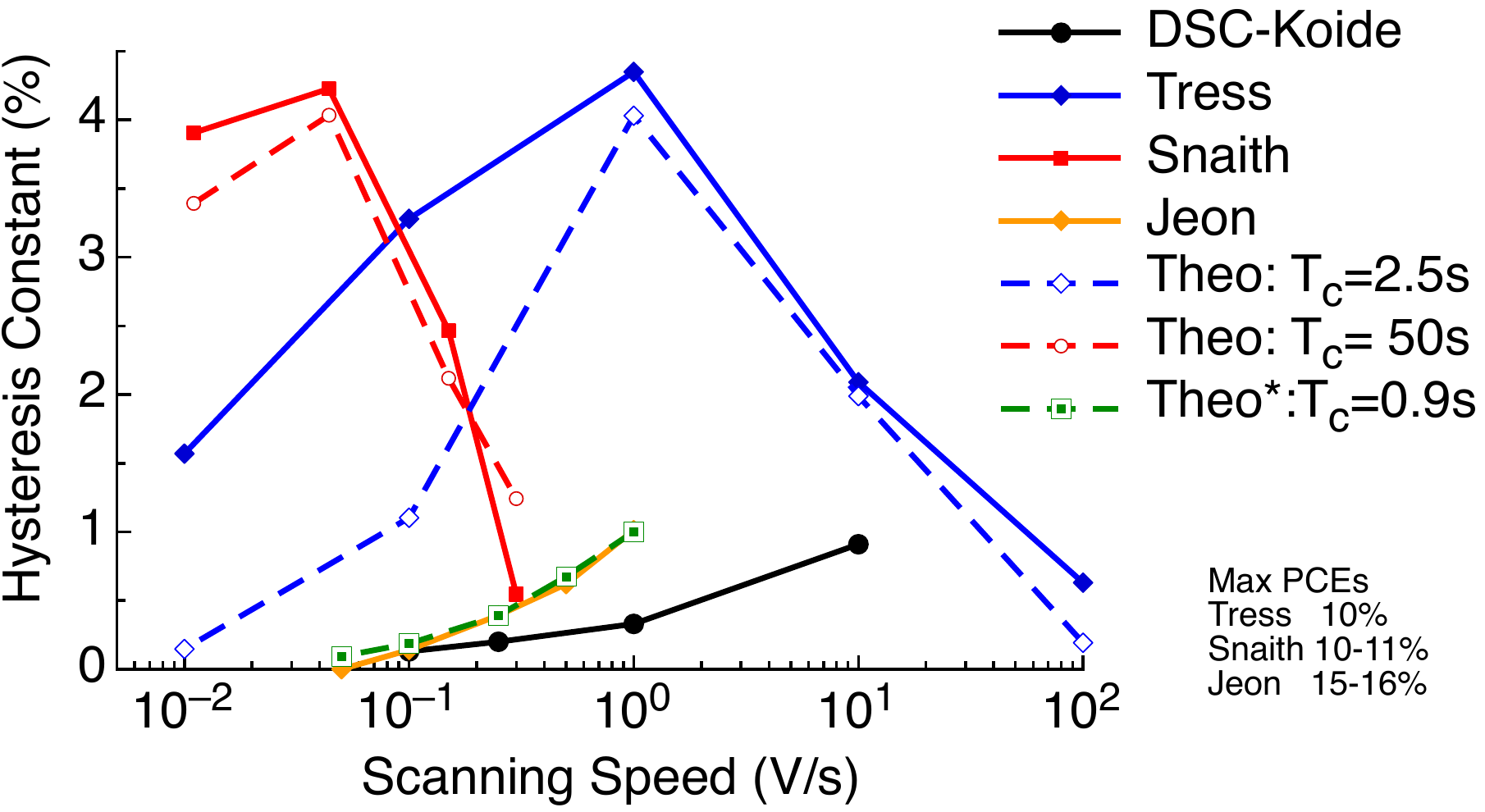}
	\caption{Scanning rate dependent hysteresis constants. The hysteresis constant is defined as the difference between the maximum PCEs of forward and reverse measurement. Experimental data is depicted by solid lines with filled points. Numerical simulation result are plotted as dashed lines with open points. T$_c$ is the polarization relaxation time used in the model.}
	\label{fig:Ehys}
\end{figure}

 What makes perovskite solar cell hysteresis mysterious is not only its large relaxation time, but also the changing of hysteresis with scan rate. For silicon solar cells and DSCs, a shorter delay time (higher scan rate) brings about more obvious hysteresis.\cite{Koide2004,Koide2005,Herman2012} In contrast, in perovskite solar cells, a shorter delay time can either induce a more or less obvious hysteresis.\cite{Unger2014,Tress2015,Snaith2014} As shown in Figure \ref{fig:Ehys}, even in the same time region, some experiments increase and others decrease. The hysteresis constant is defined as the difference between the maximum PCEs of forward and backward measurement. The presence of two peaks may reflect two different mechanisms, polarization and ion migration. A single \ce{MA+} ion takes several ps to rotate,\cite{Frost2016,Weller2015b,Leguy2015} while, the slow charge relaxation time in perovskite solar cells is in the order of 1-30 s.\cite{Unger2014,Snaith2014,Gottesman2014,Tress2015,Wei2014e} This is because the collective relaxation of millions of ions should take much longer. The timescale for a domain wall to traverse a typical device is about 0.1-1 ms.\cite{Leguy2015} For ion migration, it should take longer due to its larger energy barrier than \ce{MA+} rotation. If we select a typical 0.2 eV \cite{Haruyama2015,Azpiroz2015}energy barrier for \ce{I-} ion migration and 0.05 eV\cite{Zhou2015} for \ce{MA+} rotation, and assume the relaxation time is proportional to $e^{-E_b/kT}$, where $E_b$ is energy barrier and kT is thermal energy, the relaxation time of \ce{I-} ions is about 320 times higher than the relaxation time of \ce{MA+} ions polarization. Considering longer relaxation time for the experiment, we set capacitive charge relaxation time as 2.5 s and 250 ms for polarization in below simulations.

%\textbullet  \textbf{Tress's experiment and simulation fitting results}

\begin{figure} 
	\centering
	\includegraphics[trim={0 0mm 0 3mm},clip,width=\textwidth]{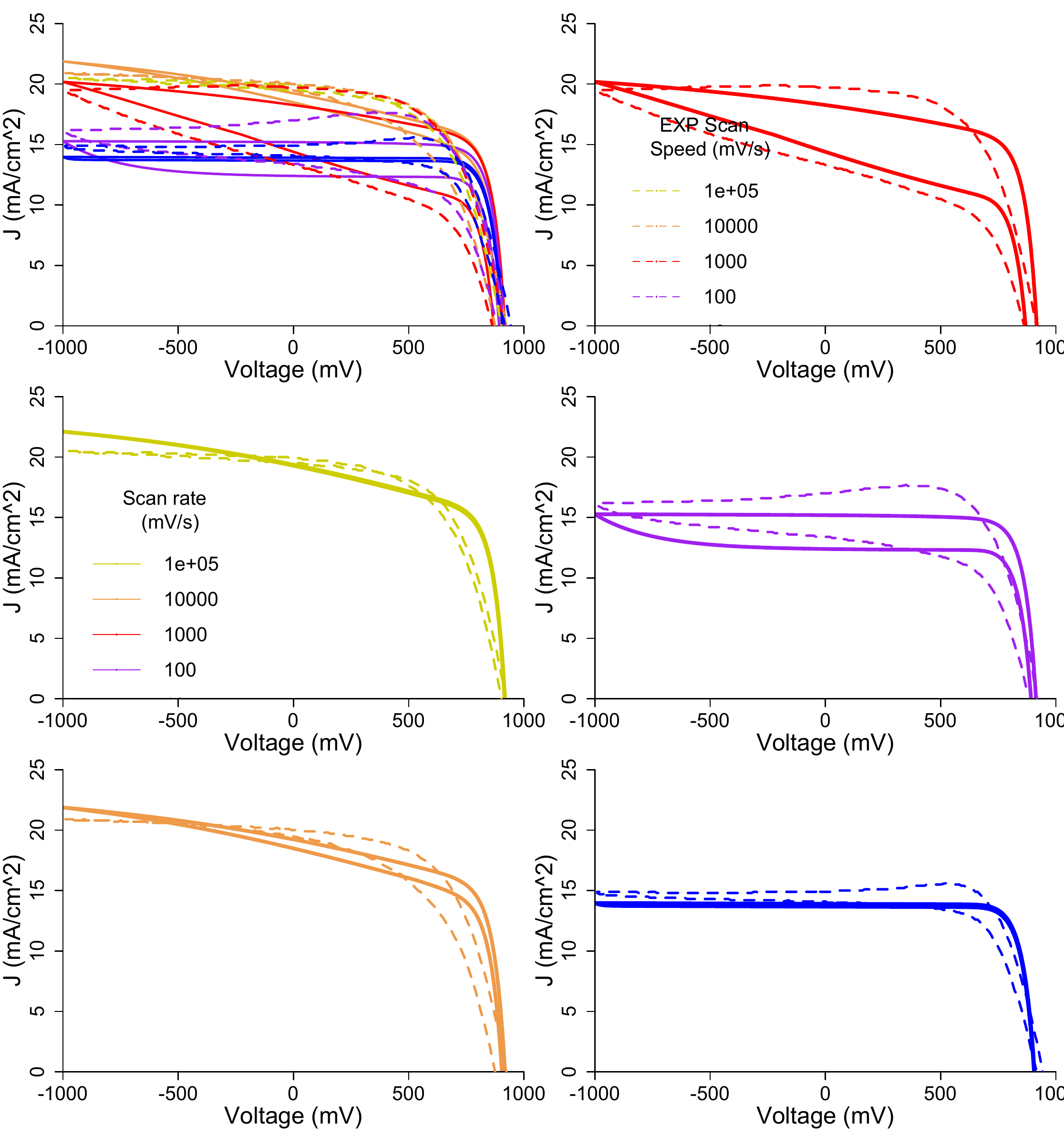} %fit_tress
	\caption{Performances of a solar cell measured forward and backward with different scan rates. Solid lines are simulation results. Dashed lines are results from Tress \textit{et al.}'s experiment. We use direct recombination without an interface.}
	\label{fig:fit-tress}
\end{figure}

Lots of experiments have been designed to study hysteresis, the most systematic experiment is the work done by Tress \textit{et al}.\cite{Tress2015} They measured one hybrid solar cell forward and backward with different scan rates from 10 to 100,000 mVs$^{-1}$. 
Every measurement begins with certain polarization states which are pre-polarized with the same condition. 
As all the measurements were taken from a single solar cell with a certain initial state we could model their J-V curves under different scan rates with one set of basic parameters. The parameters used to model the work by Tress \textit{et al.} are shown in Table \ref{tab:param}. In Figure \ref{fig:fit-tress}, experimental J-V curves are drawn with dashed lines, while simulated J-V curves are plotted with solid lines. The PCE is higher during backward scanning. The voltage falls during the backward scan, and the screening field lags behind. It follows that the screening field is higher during the backward scan than during the forward scan. That means that the hindering field is more screened during the backward scan leading to a higher efficiency.

Our simulated $V_{oc}$ are about 915 mV, which  are in good agreement with Tress's experiment. As we expect, simulated performance of the backward scan is higher than that of the forward scan. The current at voltage of -1 V with scan rate of \SI{100}{ Vs^{-1}} is \SI{22.10}{ mAcm^{-2}}, which is close to the experimental value of \SI{20.40}{ mAcm^{-2}}. At -1 V, currents modeled with scan rates of \SI{10}{ Vs^{-1}}, \SI{1}{ Vs^{-1}}, \SI{100}{ mVs^{-1}} and \SI{10}{ mVs^{-1}} are \SI{21.87}{ mAcm^{-2}}, \SI{20.18}{ mAcm^{-2}}, \SI{15.27}{ mAcm^{-2}} and \SI{13.97}{ mAcm^{-2}}, respectively. All of these current densities are close to the experiment \SI{20.90}{ mAcm^{-2}}, \SI{19.4}{ mAcm^{-2}}, \SI{16.20}{ mAcm^{-2}} and \SI{14.90}{ mAcm^{-2}} with errors smaller than \SI{1.0}{ mAcm^{-2}}. Not only do these typical values agree, but the J-V curve also have a similar shape. By applying one set of parameters for a certain solar cell, we reproduce its J-V curves under different conditions.  The agreement between Tress' experiment and our simulation suggests that our model and the proposed mechanism are correct. 

\begin{table}[]
	\caption{Assumed and experimental parameters used to simulate Tress' solar cells.}
	\label{tab:param}
	\begin{tabular}{lll}
		\hline 
		Symbol &Meaning &Value   \\
		\hline
		$E_{bgap}$&Band gap & 1.52 eV\\ 
		T &Temperature & 300 K   \\
		$I_l$    & Light intensity &1.5 AM \\
		$IPCE$& IPCE& 100\%\\ 
		$d$    & Perovskite thickness &350 nm \\
		$N_c$,$N_v$& Density of States&$3.97 \times 10^{18}cm^{-3}$.\cite{Zhou2015} \\
		$\alpha$   & Absorption coefficient & $5.7 \times10^4cm^{-1}$.\cite{Xing2013}\\ 
		A  & Voltage drop percentage & 8.4\% \\
		$D_n$  &\small  Electron diffusion coefficient & 0.030 $cm^2s^{-1}$\\
		$D_p$  &Hole diffusion coefficient& 0.063 $cm^2s^{-1}$\\
		$\tau_p$  &Polarization relaxation time & 0.25 s\\
		$\tau_c$&Capacitive charge relaxation time &  2.5 s\\
		$S_p$  &Polarization screen coefficient &50\% \\
		$S_c$&Capacitive charge screen coefficient &   49\%\\
		$\tau$&Charge carriers lifetime & 10.2 ns \\
		\hline 
	\end{tabular}
\end{table}

%\textbullet \textbf{Simulation of hysteresis constants .vs. experiment hysteresis constants}

For normal hysteresis from typical capacitive charges, as seen in Dye Sensitized Solar Cells made by Koide \textit{et al.}\cite{Koide2004,Koide2005}  it becomes more extreme when the scan rate increases, as shown in Figure \ref{fig:Ehys}. In contrast, Snaith observed anomalous hysteresis, which becomes more extreme as the scan rate is reduced. Even at extremely slow scan rates, it is still significant.\cite{Snaith2014} We interpret this phenomenon as being a result of slower ion migration. As observed in experiment, the relaxation time can be as short as 1 s,\cite{Sanchez2014} or alternatively as long as several tens of seconds.\cite{Unger2014,Snaith2014,Gottesman2014,Tress2015,Wei2014e} The performance is similar in Tress' and Snaith's experiments which suggests their solar cells have similar electronic parameters. If we change the capacitive relaxation time to 50 s without changing other parameters, our simulations give hysteresis in good agreement with Snaith's experiment.\cite{Snaith2014} 
Jeon's solar cells show higher PCEs, hence, their solar cells should have better electronic properties.\cite{Jeon2014} To model Jeon's solar cells, we increased the charge carrier lifetime to 80 ns and decreased $S_{c}$ to 10\%.  Hysteresis seen in Jeon's experiment is also repeated with a capacitive charge relaxation time of 0.9 s. All of the simulation and experimental hysteresis are shown in Figure \ref{fig:Ehys}. It is found that for very low or high scan rate measurements, the hysteresis decreases. This is due to the relaxation field cannot follow the changing of external fields for an extremely high scan rate. The screening field is constant during a forward-backward measurement. No hysteresis is expected as there is no difference between forward and backward scans. While, for the case of very slow scan, there is enough time to relax, the screening field is always proportional to the external applied field. Screening fields are the same under a certain applied voltage no mater it is a forward or a backward scan. Hence, hysteresis will not happen at very slow scan rate.

Hysteresis-free inverted perovskite solar cells  have been made from interface engineering or using \ce{PCBM} electron transfer layer.\cite{Kim2015,Miyano2016a,Hou2016} We argue these hysteresis-free devices are due to the improvement of interface states rather than the inverted structure, because normal structure perovskite solar cells have been made without large hysteresis by implementing \ce{C60}\cite{Zhao2016} or PCBM\cite{Heo2015}. As illustrated in Figure \ref{fig:mechanism}, both the polarization charge and the ionic charge are accumulated near interfaces. These charges could be compensated or neutralized by contact layers. For an ideal crystal without any defects, ions cannot migrate because there is no vacancy to go. Whereas, defects exist at interfaces, such as dangling bonds. Therefore, it is possible to reduce and even to eliminate ion migration and it induced hysteresis through interface engineering or decreasing defects in thin films. For uniform polarizations, there is no net charge overall and also no net charge in the bulk. All of the polarization charges are near surfaces or interface. In perovskite solar cells, if polarizations present, net charges should locate at interfaces between perovskite and contact layers: one end of perovskite is positive and another is negative. If we impose a contact layer with negative charges at the surface on the end with positive charges, and a contact layer with positive charges at the surface on the end with negative charges, these polarization charges are neutralized. Therefore, this part of the hysteresis also could be reduced or eliminated through interface engineering. The slow relaxed polarization of perovskite is also possible to be compensated by the polarization of contact layers. Fullerene and its derivatives, such as \ce{C60}\cite{Chaban2015} and PCBM\cite{Tada2011,Ryno2016} are polarisable. Hence they are good candidates to compensate the polarization of perovskite and eliminate hysteresis of PSCs. Some of these materials have been successfully applied to reduce or eliminate hysteresis, such as \ce{C60} in Ref.\onlinecite{Zhao2016} and PCBM in Ref.\onlinecite{Hou2016,Kim2015,Miyano2016a,Heo2015}.

% % % % % % % % % % % % % % % % % % % % % % % % % % % %
\section{Conclusion} 
%\textbullet 
%\textbf{Screening charge weaken the hindering field, promote charge transport and improve solar cells performance.}

Using numerical simulations we have confirmed that the screening effect improves perovskite solar cell performance. This improvement is more obvious in solar cells made from simple methods. On the basis of our previous DFT calculations, we argue that the screening field comes from both ion migration and polarization charge. This field weakens the hindering field, promotes charge transport and improve PCEs.

%\textbullet \textbf{Slow process induce anomalous hysteresis. The slow process could be polarization  or ion migration.}
Due to slow polarization and ion relaxation, the screening field is delayed, which leads to hysteresis. As the relaxation time of capacitive and polarization charges are in different scales, rate dependent hysteresis behaviors become more complicated. With the relaxation of capacitive and polarization charges we reproduced various measured hysteresis curves.\cite{Unger2014,Tress2015, Jeon2014} Using similar parameters and different scan rates, we also reproduced the hysteresis effects observed by Snaith \textit{et al.}. These results suggest that hysteresis is caused by two kinds of very slow relaxations. This agreement with measured hysteresis, in turn, supports our proposed mechanism that polarization and capacitive charge take part in the screening of the hindering field and improves its PCE. We show that not only can ion migration cause hysteresis, but also polarization can as well. Although, both ion migration and polarization are bulk properties, they could be affected by interface states and contact layers. Polarizable contact materials, such as fullerene and its derivatives,  may be good candidates to compensate screening fields from polarization or ionic charges and eliminate hysteresis of PSCs.

In this paper we assumed that one of the relaxations is polarization based on Gottesman's experiment,\cite{Gottesman2014} but this is not necessary. Any slow relaxation response with the ability to screen external hindering field could bring about hysteresis. This slow relaxation could also be different kinds of ion migration.  As polarization relaxation and ion migration usually exhibit similar behavior, such as thickness dependence.\cite{Kim2007,Bazant2004} To determine whether these slow responses are polarization or ion migration, more experiments and theoretical works should be carried out further. 

%The ``anomalous" hysteresis of \ce{CH3NH3PbI3} solar cell is the result of both polarization and ion migration relaxation.
\section{Acknowledgment} 
We greatly appreciate the contribution of Professor Yang Yang of UCLA and Professor Qi Chen for providing their initial data in reference \onlinecite{Zhou2014}. Their initial data which is used to plot the data ``Exp-Zhou" in Figure \ref{fig:screen} allowed us to compare the experiment and theory in detail.

% % % % % % % % % % % % % % % % % % % % % % % % % % % %
\bibliographystyle{achemso-demo}
%\bibliographystyle{apsrev4-1}
%\bibliography{aipsamp}
\bibliography{numerical_hysteresis}

\end{document}